\documentclass[aps,prl,showpacs,twocolumn,groupedaddress]{revtex4}

\usepackage{amsfonts}
\usepackage{amsmath,amssymb}
\usepackage{amsthm}
\usepackage{graphicx}
\usepackage[dvips]{color}

\renewcommand{\d}{{\rm d}}
\newcommand{\e}{{\rm e}}
\newcommand{\imai}{{\rm i}}

\newcommand{\pabl}[2]{\frac{\partial{#1}}{\partial{#2}}}

\begin{document}


\title{Coherence and Spatial Resolution of Transport in Quantum
  Cascade Lasers}

\author{Andreas Wacker}\email{Andreas.Wacker@fysik.lu.se}

\affiliation{Mathematical Physics, Lund University, Box 118, 22100 Lund, Sweden}
\date{\today, submitted to Proceedings of HCIS-15 (July 2007), physica status
  solidi (c)}


\begin{abstract}The method of nonequilibrium Greens functions allows for a
spatial and energetical resolution of the electron current in
Quantum Cascade Lasers. While scattering does not change the 
spatial position of carriers, the entire spatial evolution of charge
can be attributed to coherent transport by complex wave functions.
We discuss the hierarchy of transport models and derive the density
matrix equations as well as the hopping model starting from the
nonequilibrium Greens functions approach.
\end{abstract}

\maketitle

\section{Introduction}

Since the first realization in 1994 \cite{FaistScience1994} Quantum
Cascade Lasers (QCLs) have become an important tool for
IR-spectroscopy. The extension towards the THz region
\cite{KohlerNature2002} has opened up possibilities for a variety of
applications in rapid-but-precise hazardous chemical sensing, concealed
weapon detection, non-invasive medical and biological diagnostics, and
high-speed telecommunications \cite{LeeScience2007}.

The operation of QCLs is based on electronic transitions between
different subbands within the conduction  band of a semiconductor
heterostructure.  Using a sophisticated sequence of wells and
barriers, the electrons are guided into the upper laser level at the
operating bias, thus creating population inversion for a pair of levels in the
active region. By modifying the layer thicknesses, the transition energy
can be varied in a large range and operating lasers with wavelengths
between 2.95$\mu$m \cite{DevensonAPL2007} and 217 $\mu$m (1.39 THz)
\cite{ScalariAPL2006} have been realized within the last year.  This
covers a large part of the electromagnetic spectrum from the
near-infrared to the proximity of fast electrical circuits (albeit
there is a gap at the Reststrahlenband).

Conventionally, QCLs are modeled by rate equation schemes either for
the average electron densities in the subbands
\cite{CapassoJMP1996,HarrisonAPL1999,IndjinJAP2002},  or the
occupation of the individual states
\cite{IottiPRL2001,CallebautAPL2004,BonnoJAP2005,JirauschekJAP2007,GaoAPL2007}.
The latter ones are often simulated with the Monte-Carlo technique,
which suggests this denomination, albeit the term {\em hopping
transport} \cite{TsuPRB1975,CaleckiJP1984} seems to be more
appropriate for this type of models. Such simulations allowed for a
continuous improvement of device performance by optimizing the layer
structure for an appropriate ratio of scattering matrix elements and
resonance conditions. Hopping or rate equation models can however 
not describe coherent
effects, which are of some relevance for the tunneling transition
between the injector into the upper laser level
\cite{SirtoriIEEE1998,CallebautJAP2005}. Furthermore, the broadening
of the gain transition can only be qualitatively estimated within
such models. To overcome these limitations, a quantum transport model
based on nonequilibrium Green functions (NEGF) was developed
\cite{LeePRB2002}. It was demonstrated that the microscopic current
flow is due to coherent evolution of wave packets rather than the
spatial translation by scattering transitions \cite{LeePRB2006}. Here
this idea is further elaborated with a particular focus on the
relation between the NEGF  model, 
density matrix equations \cite{IottiPRB2005}, and the above mentioned
hopping transport models.

The paper is organized as follows: In Section 2, the different
concepts for calculating a current in QCLs (or similar semiconductor
heterostructures elements) are discussed. A key result is that the
entire current is carried by nondiagonal elements of the density
matrix (coherences), as also discussed in Ref.~\cite{IottiPRB2005}.
In Section 3, we present numerical examples for the different
representations of current. In the more technical sections 4 and 5
it is shown how the density matrix equations, and the standard 
hopping models are derived by successive simplifications of
the Greens function technique.

\section{Modeling the current}

In planar semiconductor heterostructures, such as QCLs or
superlattices, it is appropriate to use a set of normalized basis states
$\frac{1}{\sqrt{A}}\varphi_{\alpha}(z)\e^{\imai {\bf k}
\cdot {\bf r}}$ which separate the
behavior in growth direction ($z$) with quantum number $\alpha$ from
the plane wave behavior (${\bf k}$) in the  $(x,y)$-plane of total area
  $A$.
The Hamilton operator is
written as 
\begin{equation}
\hat{H}=\underbrace{\hat{H}^0}_{=\hat{\vec{p}}\frac{1}{2m(z)}\hat{\vec{p}}
+V_c(z)+e\phi(z)}
+\hat{H}^{\textrm{scatt}}
\label{EqHam}
\end{equation}
where $V_c(z)=V_c(z+d)$ is the conduction band edge for our structure
with period $d$. $\phi(z)$ is the (self-consistent) electric potential
satisfying $\phi(z+d)=\phi(z)-Fd$, where $F$ is the average electric
field along the structure. 
Here it is important to note, that $\hat{H}^0$ is diagonal in
${\bf k}$ due to the translational symmetry of the perfect QCL
structure in the $(x,y)$-plane. In contrast, impurities, phonons and
possibly the presence of other electrons constitute scattering terms
of the form $\hat{a}_{\alpha {\bf k}}^{\dag}\hat{a}_{\beta {\bf k}'}$ 
with ${\bf k}\neq {\bf k}'$ in
$\hat{H}^{\textrm{scatt}}$. Here $\hat{a}_{\alpha {\bf k}}^{\dag}$ and
$\hat{a}_{\alpha {\bf k}}$ are the standard creation and annihilation
operators in occupation number representation, respectively.
In the following we use the Wannier basis for our calculations, 
see \cite{WackerPR2002}, which provides a periodic array of states
satisfying $\varphi_{\alpha}(z+d)=\varphi_{\alpha'}(z)$.
The same property holds for Wannier-Stark (WS) states as well, which in
addition diagonalize $\hat{H}^0$ with
energies $E_{\alpha}(k)$.

The current can be evaluated in two ways: The current density averaged
over the entire sample reads
\begin{equation}\begin{split}
J=&\frac{e}{V}\left\langle \frac{\d}{\d t}\hat{z}\right\rangle=
\frac{\imai e}{\hbar V}\langle [\hat{H}^0,\hat{z}]\rangle\\
=&
\frac{2\mbox{(for spin)}\imai e}{\hbar V} \sum_{\alpha,\beta,{\bf k}}
W_{\beta\alpha}
\rho_{\alpha\beta}({\bf k})
\label{EqJaverage}
\end{split}\end{equation}
where $\rho_{\alpha\beta}({\bf k})=\langle
\hat{a}_{\beta {\bf k}}^{\dag}\hat{a}_{\alpha {\bf k}}\rangle$ 
is the density matrix
(here defined to be diagonal in ${\bf k}$), 
$W_{\beta\alpha}=\sum_{\gamma}H^0_{\beta\gamma}z_{\gamma\alpha}-
z_{\beta\gamma}H^0_{\gamma\alpha}$,
and $V=NdA$ is the normalization volume of the system with $N$
periods. Note that the second contribution of the Hamiltonian
(\ref{EqHam}) provides $\langle [\hat{H}^{\rm scatt},\hat{z}]\rangle=0$, 
as the operator $\hat{H}^{\rm scatt}$ is only a 
function of $\hat{r}$, but not of $\hat{p}$, for
all scattering processes typically considered \cite{LeePRB2006}.

The local current density is given by
\begin{equation}\begin{split}
J(z)
=&\frac{2\mbox{(for spin)}}{A}\int\d x\d y\, 
\frac{e}{2m}\\
\times& \left\langle \hat{\Psi}^{\dag}(\vec{r})
\frac{\hbar}{\imai}\pabl{}{z}\hat{\Psi}(\vec{r})
+\left(
\frac{\hbar}{\imai}\pabl{}{z}\hat{\Psi}(\vec{r})\right)^{\dag}
\hat{\Psi}(\vec{r})\right\rangle 
\\
=&\frac{e}{mA}
\sum_{\alpha \beta {\bf k}}
\langle\hat{a}_{\alpha {\bf k}}^\dag\hat{a}_{\beta {\bf k}}\rangle\\
\times&
\left\{\varphi_{\alpha}^*(z)\frac{\hbar}{\imai }
\varphi'_{\beta}(z)
+\left[\frac{\hbar}{\imai }
\varphi'_{\alpha}(z)\right]^*
\varphi_{\beta}(z)\right\}
\label{EqJz}
\end{split}\end{equation}
where the expansion $\hat{\Psi}(\vec{r})=
\sum_{\beta{\bf k}}\varphi_{\beta}(z)\e^{\imai{\bf k}\cdot{\bf r}}
\hat{a}_{\beta k}/\sqrt{A}$ for the field operators
was used. 
Averaging over $z$ and  using the commutator relation
\begin{equation}
[\hat{H}^0,\hat{z}]
=-\frac{\hbar^2}{2}
\left(\frac{1}{m(z)}\pabl{}{z}+\pabl{}{z}\frac{1}{m(z)}\right)
\label{EqCommRelation}
\end{equation}
provides directly 
Eq.~(\ref{EqJaverage}).

It is important to notice, that the 
matrix  $W_{\alpha\beta}$ is anti-hermitian. Thus the diagonal elements
vanish for a set of real basis functions and consequently the entire
current is due to the nondiagonal elements of the density matrix
$\rho_{\alpha\beta}({\bf k})$. The same holds for Eq.~(\ref{EqJz}):
If the wave functions $\varphi_{\alpha}(z)$ are real, the 
diagonal elements of the density matrix do not provide any
contribution to the current.
Now both the Wannier and Wannier-Stark basis functions can be chosen
real and therefore in both cases
the current is entirely being carried by the non-diagonal elements
of the density matrix $\rho_{\alpha\beta}({\bf k})$.

Working with NEGF \cite{HaugJauhoBook1996}, 
the density matrix is given by
\begin{equation}
\rho_{\alpha\beta}({\bf k})=\int \frac{\d E}{2\pi\imai} 
G_{\alpha\beta}^<({\bf  k},E)\, .
\end{equation}
Thus, the correlation functions $G^<(E)$ can be viewed as the
energy-resolved density matrix and Eq.~(\ref{EqJz})
can be generalized to the energy-resolved current density 
\begin{equation}\begin{split}
J(E,z)
&=\frac{e}{mA} \sum_{\alpha \beta {\bf k}}
\frac{1}{2\pi\imai} G_{\beta\alpha}^<({\bf k},E)\\
&\times
\left\{\varphi_{\alpha}^*(z)\frac{\hbar}{\imai }
\varphi'_{\beta}(z)
+\left[\frac{\hbar}{\imai }
\varphi'_{\alpha}(z)\right]^*
\varphi_{\beta}(z)\right\}
\label{EqJEz}
\end{split}\end{equation}

This equation becomes of particular interest, if one considers a
special basis set
of states  $\Psi_{n{\bf k}}(E,z)$, 
which diagonalize $G_{\beta\alpha}^<({\bf k},E)/(2\pi\imai)$ with 
the real (and positive) eigenvalues $f_{n{\bf k}}(E)$. Then the current 
as well as the density is represented by an incoherent superposition of
complex wave functions  $\Psi_{n{\bf k}}(E,z)$ at each energy.
The fact that these wave functions carry the entire current manifests
the coherent nature of current evolution in QCLs as well as 
related structures such as superlattices.

\section{Numerical examples}

\begin{figure}[tb]
\includegraphics[width=\linewidth]{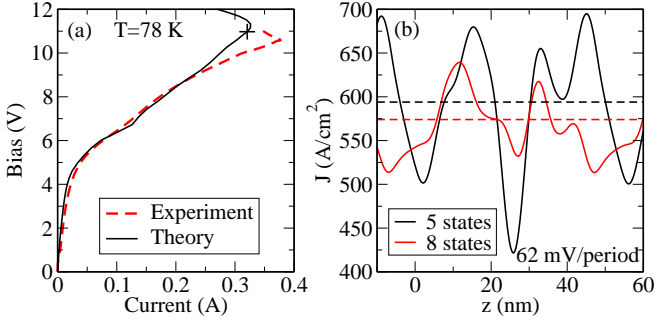}
\caption{(a) Current-voltage characteristic for the 
THz-QCL of ref.~\cite{KumarAPL2004}. In the calculations (full line) 
the bias was
taken as 177 times the voltage drop per period and the area as 
$A=54 \cdot 10^{-5}\textrm{cm}^2$. The cross marks the operation point
at a bias drop of 62 mV per period. The experimental data (dashed
line) are by courtesy of S. Kumar. (b) Spatially resolved current
density from Eq.~(\ref{EqJz}) for a calculation with 5 and 8 Wannier
states per period, respectively (full lines). The dashed lines give the
result from Eq.~(\ref{EqJaverage}) for comparison.}
\label{Figcurrent}
\end{figure}

Let's consider the THz-QCL from Kumar {\it et al.}\cite{KumarAPL2004},
which operates above 77 K. The current-voltage characteristic
evaluated via Eq.~(\ref{EqJaverage}) is shown in
Fig.~\ref{Figcurrent}(a) and good quantitative agreement with the
experimental data is found. (Details of the calculation are given in
\cite{BanitAPL2005}.) In Fig.~\ref{Figcurrent}(b) this current
(dashed line) is compared with the local current density (full line)
evaluated via Eq.~(\ref{EqJz}). While current continuity requires a
constant $J(z)$ in the stationary case, the evaluated local current
exhibits spatial oscillations. The amplitude of these oscillations
decreases with the number of Wannier states per period employed in the
calculations. This suggests that this artificial effect is due to the
lack of completeness if only a finite number of basis states is taken
into account for. The result from Eq.~(\ref{EqJaverage}) corresponds
to the spatial average and is far less sensitive to the number of
states employed. This shows that the current evaluated by
Eq.~(\ref{EqJz}) has to be taken with care and Eq.~(\ref{EqJaverage})
is preferable.

\begin{figure}[tb]
\includegraphics[width=\linewidth]{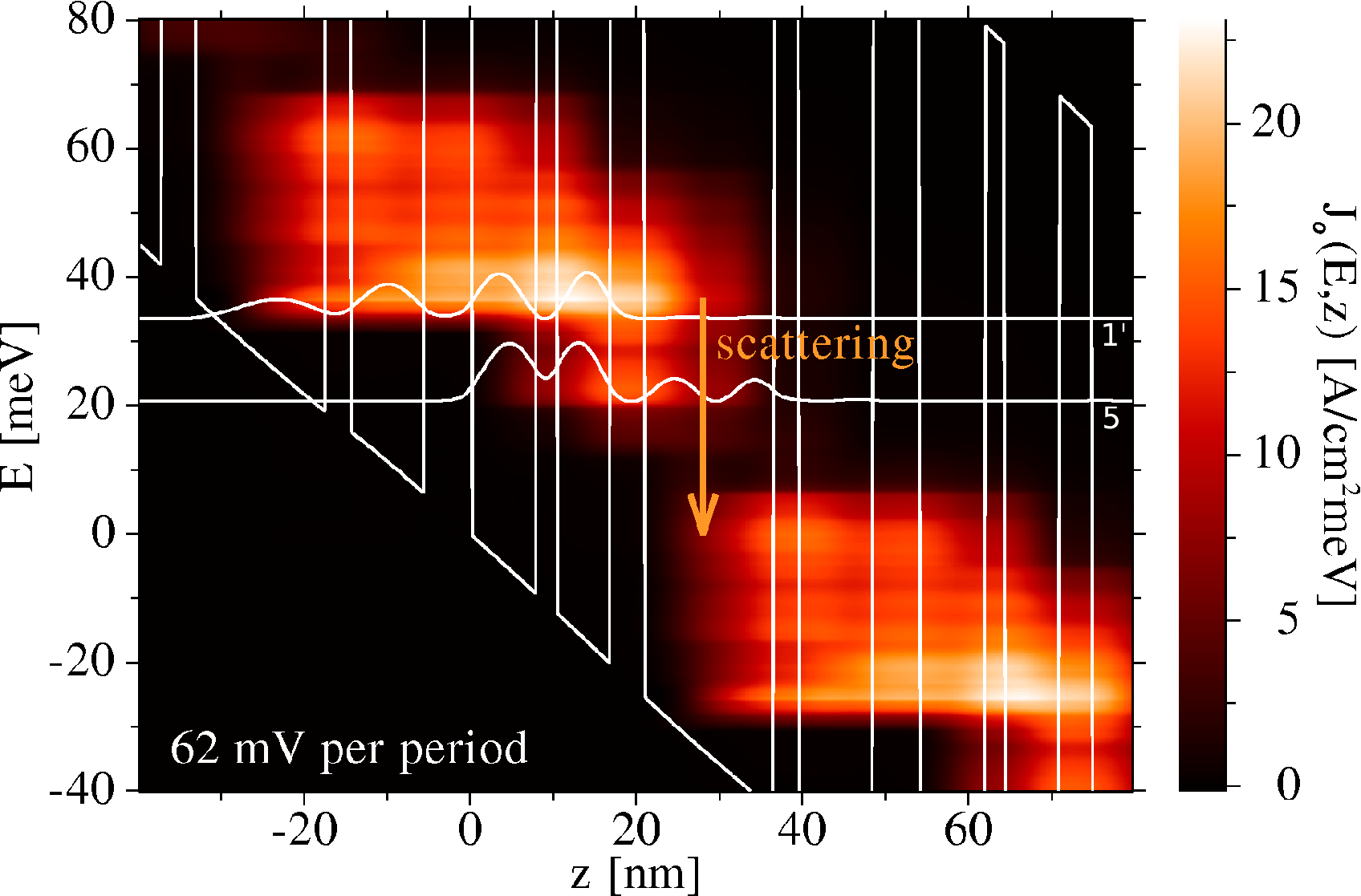}\\
\includegraphics[width=\linewidth]{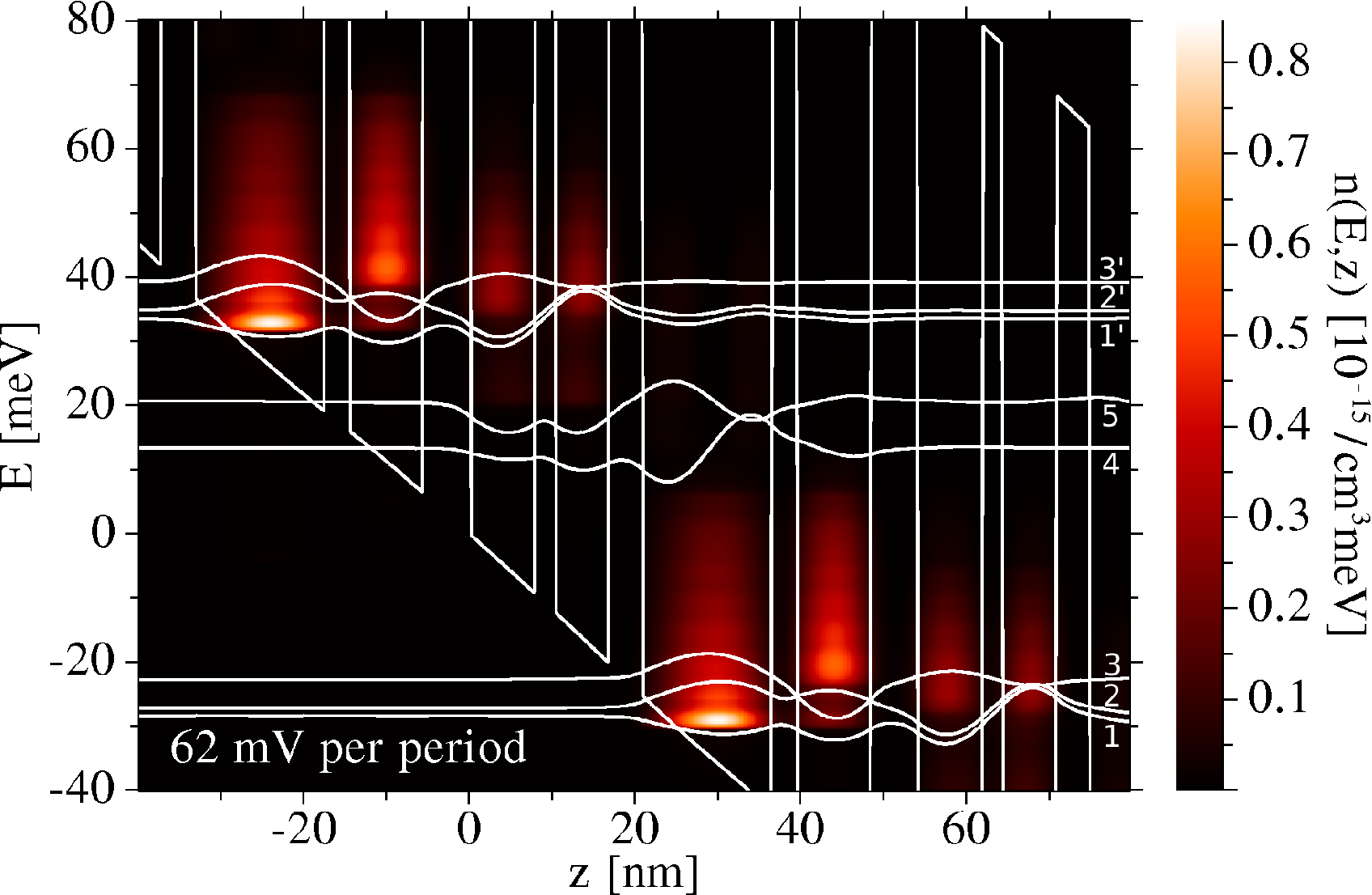}
\caption{Upper panel panel: Spatially and energetically 
resolved current density evaluated by Eq.~(\ref{EqJEz}). 
The WS states $\varphi_\alpha^2(z)$ corresponding to 
the upper (1') and lower (5) laser level are depicted for orientation. 
The vertical array marks a representative
  scattering transitions. Lower panel:
Spatially and energetically resolved particle density
  evaluated by Eq.~(\ref{EqnEz}). The lowest five WS states $\varphi_\alpha(z)$
are displayed.}\label{FigCurrentEZ}
\end{figure}

In the upper panel of 
Fig.~\ref{FigCurrentEZ} the energetically resolved current density
from Eq.~(\ref{EqJEz}) is displayed. At each energy one observes a
current flow in $z$ direction due to the presence of nondiagonal
elements in $G_{\beta\alpha}^<({\bf k},E)$. In order to satisfy the
continuity of current, scattering transitions transfer particles
between different energies, where the coherent evolution of the 
current continues. In addition, there are also elastic
scattering events, where the current continues at the same energy, but
with a different parallel momentum  ${\bf k}$, which are visible
in corresponding ${\bf k}$-resolved plots.

For comparison the energetically resolved particle density 
(see also \cite{KubisJCE2007})
\begin{equation}
n(E,z)
=\frac{2}{A} \sum_{\alpha \beta {\bf k}}
\frac{1}{2\pi\imai}G_{\beta\alpha}^<({\bf k},E)
\varphi_{\alpha}^*(z)\varphi_{\beta}(z)
\label{EqnEz}
\end{equation}
is shown in the lower panel of 
Fig.~\ref{FigCurrentEZ}. It is intriguing to see, that neither the
density nor the current profile follow the spatial profile of the 
WS-states. Furthermore note, that the states 1 and 2 resemble the
binding and anti-binding combination of two more localized states.

\begin{figure}[tb]
\includegraphics[width=\linewidth]{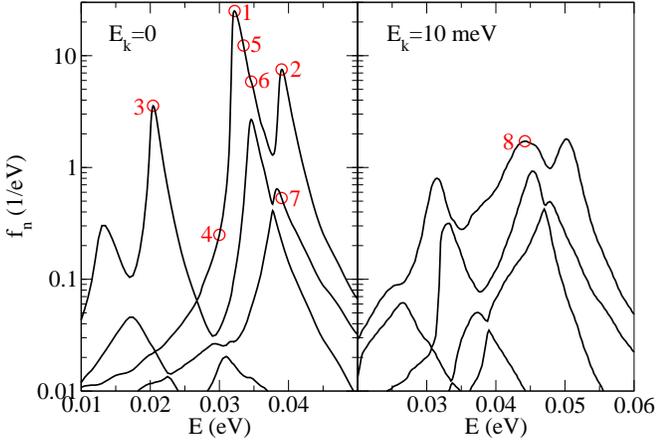}
\caption{Eigenvalues of the matrix  $G_{\beta\alpha}^<({\bf k},E)/(2\pi\imai)$
as a function of energy for two different values of ${\bf k}$. The states
corresponding to the eigenvalues denoted by circles are displayed in 
Fig.~\ref{FigWave}.
}
\label{FigFvalues}
\end{figure}

In Fig.~\ref{FigFvalues} the eigenvalues of 
$G_{\beta\alpha}^<({\bf k},E)/(2\pi\imai)$ are displayed as a
function of energy. One can identify distinct peaks indicating the
presence of broadened quasiparticle states. These can be attributed
to specific branches in the eigenvalue spectrum, which however 
mix with each other at crossing points.
The essential structure for ${\bf k}=0$ is repeated for
finite {\bf k}-values with a shift in energy by $E_k=\hbar^2k^2/2m$.
For $E_k=10$ meV the width of the peaks is larger 
as more scattering states are present than for $E_k=0$ meV.

\begin{figure}[tb]
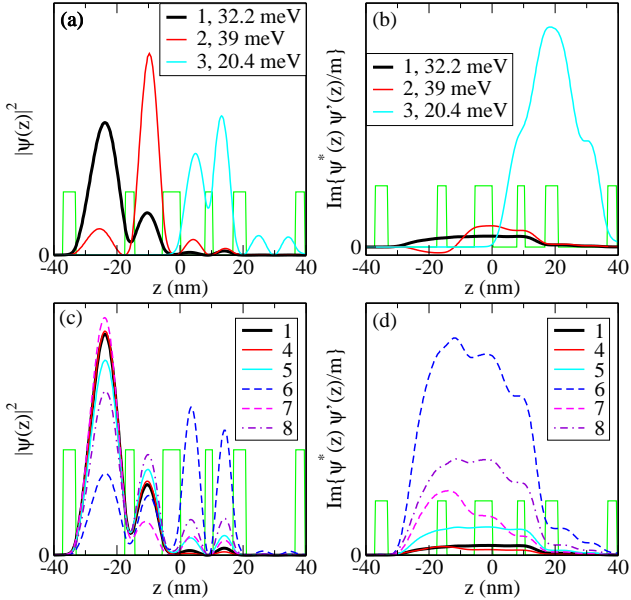

\includegraphics[width=0.48\linewidth]{figure4a.eps}
\includegraphics[width=0.48\linewidth]{figure4b.eps}\\
\includegraphics[width=0.48\linewidth]{figure4c.eps}
\includegraphics[width=0.48\linewidth]{figure4d.eps}
\caption{Wave functions
$\Psi_{n{\bf k}}(E,z)$ which diagonalize 
$G_{\beta\alpha}^<({\bf k},E)/(2\pi\imai)$ for different energies and
${\bf k}$ corresponding to the eigenvalues denoted by  the 
circles in Fig.~\ref{FigFvalues}.}
\label{FigWave}
\end{figure}

In Fig.~\ref{FigWave}(a,b) the wave functions corresponding to the three
largest eigenvalue peaks are shown. 
They describe the spatial structure of both the
electron density and current 
displayed in Fig.~\ref{FigCurrentEZ}. Thus they give a better description of
the ongoing behavior than Wannier or Wannier-Stark states.
In many cases these states are essentially unchanged 
if one follows a single branch of
eigenvalues in Fig.~\ref{FigFvalues}. E.g., the states 1 and 4 differ 
only very slightly, see Fig.~\ref{FigWave}(c,d). However a strong energy
dependence can occur due to mixing
effects if different branches of eigenvalues come close to each other or even
cross. This can be seen in the sequence for states 5, 6, and 7. 
The states for finite ${\bf k}$ are related to the corresponding states at 
${\bf k}=0$ at the lower energy $E-E_k$. 
E.g., state 8  corresponds to states 5,6 
(which are about 10 meV lower in energy $E$),
albeit the mixing between different branches makes a detailed comparison 
difficult.

\section{Density-matrix equations}

Now we want to study the relation between the different approaches.
In the NEGF approach, the results are determined by the
lesser Greens function. For the stationary state, where
\[
G^{<}(t,t')=\int\frac{\d E}{2\pi}G^{<}(E)\e^{-\imai(t-t')E/\hbar}\, ,
\]
Eq.~(5.4) of Ref.~\cite{HaugJauhoBook1996} provides us with
\begin{equation}
\begin{split} 
&\sum_{\gamma}G^<_{\alpha\gamma}(E,{\bf k}) H^0_{\gamma \beta}({\bf k})-
H^0_{\alpha\gamma}({\bf k})G^<_{\gamma\beta}(E,{\bf k})\\
&=\sum_{\gamma}\Big[ 
\Sigma^{\rm ret}_{\alpha\gamma}(E,{\bf k})G^{<}_{\gamma\beta}(E,{\bf k}) 
-G^{\rm ret}_{\alpha\gamma}(E,{\bf k})\Sigma^{<}_{\gamma\beta}(E,{\bf k})\\
&+\Sigma^{<}_{\alpha\gamma}(E,{\bf k})G^{\rm adv}_{\gamma\beta}(E,{\bf k})
-G^{<}_{\alpha\gamma}(E,{\bf k})\Sigma^{\rm adv}_{\gamma\beta}(E,{\bf k}) 
\Big]
\label{EqGless}
\end{split}\end{equation}
The self-energies are evaluated in the self-consistent
Born-approximation providing
\begin{equation}\begin{split} 
\Sigma^{<}_{\alpha\beta}&(E,{\bf k})=\sum_{\gamma\delta {\bf q}}
V_{\alpha\gamma}({\bf q})V_{\delta\beta}(-{\bf q})\\
\times&
\big[n_{\bf q} G^{<}_{\gamma\delta}(E-\hbar\omega_{\bf q},{\bf k}-{\bf q})\\
&+(n_{-{\bf q}}+1) G^{<}_{\gamma\delta}(E+\hbar\omega_{-{\bf q}},{\bf k}-{\bf q})\big]
\\
\Sigma^{\rm ret/adv}_{\alpha\beta}&(E,{\bf k})=\sum_{\gamma\delta {\bf q}}
V_{\alpha\gamma}({\bf q})V_{\delta\beta}(-{\bf q})\\
\times&
\big[(n_{\bf q}+1) G^{\rm ret/adv}_{\gamma\delta}(E-\hbar\omega_{\bf q},{\bf k}-{\bf q})
\\
&+n_{-{\bf q}}G^{\rm ret/adv}_{\gamma\delta}(E+\hbar\omega_{-{\bf
    q}},{\bf k}-{\bf q})\big]
\label{EqSE}
\end{split}\end{equation}
For illustrative purpose only phonon scattering with a single lateral mode
${\bf q}$ is taken into account here and the nondegenerate case is considered
(otherwise additional terms with $G^<$ appear in
$\Sigma^{\rm adv/ret}$). However, neither of these simplifications
was performed in the numerical examples discussed above.

Neglecting any broadening effects, the full Greens functions can be
approximated by the bare Greens functions
\begin{equation}
\begin{split}
G^{\rm ret/adv}_{\alpha\beta}(E,{\bf k})\approx&\delta_{\alpha\beta}
\frac{1}{E-E_{\beta}({\bf k})\pm\imai 0^+}\\
G^{<}_{\alpha\beta}(E,{\bf k})\approx&2\pi\imai
\rho_{\alpha\beta}\delta(E-E_{\alpha\beta}({\bf k}))\label{EqGbare}
\end{split}\end{equation}
A key issue is that we allow for a nondiagonal density matrix, which
makes it difficult to address a specific energy $E_{\alpha\beta}({\bf k})$ 
to the respective $\delta$-function. A first guess is that 
$E_{\alpha\beta}({\bf k})$ is somehow related to  
$E_{\alpha}({\bf k})$ and/or $E_{\beta}({\bf k})$.

Now Eq.~(\ref{EqSE}) is inserted into  Eq.~(\ref{EqGless}) and
subsequently, the approximations (\ref{EqGbare})
are inserted in the right-hand side.
Integrating over $E$ and dividing by $2\pi\imai$, provides 
\begin{equation}
\begin{split} 
&\sum_{\gamma}\rho_{\alpha\gamma}({\bf k}) H^0_{\gamma \beta}({\bf k})-
H^0_{\alpha\gamma}({\bf k})\rho_{\gamma\beta}({\bf k})\\
&=\sum_{\gamma\delta {\bf q}}\Big[ 
\frac{n_{-{\bf q}}V_{\alpha\delta}({\bf q})V_{\delta\gamma}(-{\bf q})
\rho_{\gamma\beta}({\bf k})}
{E_{\gamma\beta}({\bf k})-E_{\delta}({\bf k}-{\bf q})
+\hbar\omega_{-{\bf q}}+\imai 0^+}\\
&\quad -\frac{n_{\bf q}V_{\alpha\gamma}({\bf q})
\rho_{\gamma\delta}({\bf k}-{\bf q})V_{\delta\beta}(-{\bf q})}
{E_{\gamma\delta}({\bf k}-{\bf q})-E_{\alpha}({\bf k})
+\hbar\omega_{\bf q}+\imai 0^+}\\
&\quad +\frac{n_{\bf q}V_{\alpha\delta}({\bf q})
\rho_{\delta\gamma}({\bf k}-{\bf q})V_{\gamma\beta}(-{\bf q})}
{E_{\delta\gamma}({\bf k}-{\bf q})-E_{\beta}({\bf k})
+\hbar\omega_{\bf q}-\imai 0^+}\\
&\quad -\frac{n_{-{\bf q}}\rho_{\alpha\gamma}({\bf k})
V_{\gamma\delta}({\bf q})V_{\delta\beta}(-{\bf q})}
{E_{\alpha\gamma}({\bf k})-E_{\delta}({\bf k}-{\bf q})
+\hbar\omega_{-{\bf q}}-\imai 0^+}\Big]\\
&+\textrm{terms with } n_{\bf q}\to n_{-{\bf q}}+1 \textrm{ and } 
\hbar\omega_{\bf q}\to -\hbar\omega_{-{\bf q}} 
\label{EqDenstityMatrix}
\end{split}\end{equation}

Setting $E_{\gamma\beta}({\bf k})=E_{\gamma}({\bf k})$,
$E_{\gamma\delta}({\bf k}-{\bf q})=E_{\gamma}({\bf k}-{\bf q})$,
$E_{\delta\gamma}({\bf k}-{\bf q})=E_{\gamma}({\bf k}-{\bf q})$, and
$E_{\alpha\gamma}({\bf k})=E_{\gamma}({\bf k})$ in the subsequent
lines on the right-hand side, one obtains precisely the 
density-matrix kinetics of Sec IID of \cite{IottiPRB2005}
in the so called complete collision limit. In this kinetics, the 
left-hand side has the additional term
$\imai\hbar\tfrac{\d \rho_{\alpha\beta}({\bf k})}{\d t}$, 
which however vanishes in the stationary case considered here.

In the density matrix equations, the choice of 
$E_{\alpha\beta}({\bf  k})$, which accompanies the density matrix 
$\rho_{\alpha\beta}({\bf k})$ on the right-hand side, can be related
to the way, the Markov limit is performed. Here different choices have
been suggested\cite{RossiPreprint2007,PedersenPRB2007}, which is 
an issue of ongoing debate. However, as shown below, the nondiagonal
density matrices are small unless 
$|E_{\beta}({\bf k})-E_{\alpha}({\bf k})|\lesssim \Gamma$. If the
properties of the system are constant on this energy scale, e.g., the
temperature is larger than $\Gamma/k_B$, the specific choice of
$E_{\alpha\beta}({\bf  k})$ within the energy interval 
$[E_{\beta}({\bf k}),E_{\alpha}({\bf k})]$ is not
of central relevance. Thus, the results for different choices should
not differ dramatically as observed in \cite{PedersenPRB2007}.
In the opposite case of small temperature ($<\Gamma/k_B$), broadening
effects become of importance, which renders the density matrix
approach questionable anyway.

\section{Hopping model}
The ambiguity of choosing $E_{\alpha\beta}({\bf k})$ vanishes, if we assume
that the diagonal density matrices  
$\rho_{\beta\beta}({\bf k})=f_{\beta}({\bf k})$
dominate the scattering terms which constitute
the right-hand side of Eq.~(\ref{EqDenstityMatrix}). This
makes particular sense, if the states are chosen as 
the eigenstates of $\hat{H}^0$.
For $\alpha=\beta$ we find
\begin{equation}
\begin{split} 
0=&- 2\pi\imai \sum_{\delta {\bf q}}
n_{-{\bf q}}|V_{\alpha\delta}({\bf q})|^2\\
&\quad\times \delta(E_{\alpha}({\bf k})-E_{\delta}({\bf k}-{\bf q})
+\hbar\omega_{-{\bf q}})f_{\alpha}({\bf k})\\
&+ 2\pi\imai \sum_{\gamma {\bf q}}
n_{\bf q}|V_{\alpha\gamma}({\bf q})|^2\\
&\quad  \times\delta(E_{\gamma}({\bf k}-{\bf q})-E_{\alpha}({\bf k})
+\hbar\omega_{\bf q})f_{\gamma}({\bf k}-{\bf q})\\
+& \textrm{terms with } n_{\bf q}\to n_{-{\bf q}}+1 \textrm{ and } 
\hbar\omega_{\bf q}\to -\hbar\omega_{-{\bf q}} 
\end{split}\end{equation}
This is just the difference of out-scattering and in-scattering 
transition rates for the state $(\alpha,{\bf k})$, where the 
scattering rates are evaluated by
Fermi's golden rule. This defines the hopping model
\cite{TsuPRB1975} which
has been frequently applied to QCLs 
\cite{IottiPRL2001,CallebautAPL2004,BonnoJAP2005,JirauschekJAP2007,GaoAPL2007}.
It is usually solved by the Monte Carlo technique and 
provides the stationary occupations $f_{\alpha}({\bf k})$.

For $\alpha\neq\beta$ the left-hand side of
Eq.~(\ref{EqDenstityMatrix}) 
provides the term $[E_{\beta}({\bf k})-E_{\alpha}({\bf k})]
\rho_{\alpha\beta}$ in the eigenstate basis. Again the right-hand side
has the magnitude of $\Gamma\times {\cal O}\{f_\alpha\}$, where 
$\Gamma/\hbar$ is the magnitude of the scattering rate for a single level.
Therefore does the assumption, that the diagonal elements dominate the
density matrix, become questionable if a pair of levels satisfies  
$|E_{\beta}({\bf k})-E_{\alpha}({\bf k})|\lesssim \Gamma$ which is typical
for level crossings, see also the discussion in \cite{CallebautJAP2005}.

The evaluation of the current is a subtle issue, as the current
is entirely contained in the nondiagonal density matrices as discussed
above. Now Eq.~(\ref{EqJaverage}) gives in the eigenstate basis:
\begin{equation}
J=-\frac{e}{\hbar V} \sum_{\alpha\neq \beta,{\bf k}}
z_{\beta\alpha}\Im\left\{[E_{\beta}({\bf k})-E_{\alpha}({\bf k})]
\rho_{\alpha\beta}\right\}
\end{equation}
where the antisymmetry of $W_{\alpha\beta}$ was used. Now
$[E_{\beta}({\bf k})-E_{\alpha}({\bf k})]
\rho_{\alpha\beta}$ is precisely the left-hand side of 
Eq.~(\ref{EqDenstityMatrix}) and 
restricting to the dominating 
diagonal density matrices on the right-hand side we obtain
\begin{equation}\begin{split}
J=&\frac{2\pi e}{\hbar V} \sum_{\alpha\neq\beta,{\bf k}}
z_{\beta\alpha}\\
\times& \Big[\sum_{\delta {\bf q}}
n_{-{\bf q}}V_{\alpha\delta}({\bf q})V_{\delta\beta}(-{\bf q})\\
&\times \delta(E_{\beta}({\bf k})-E_{\delta}({\bf k}-{\bf q})
+\hbar\omega_{-{\bf q}})f_{\beta}({\bf k})\\
&-\sum_{\gamma {\bf q}}n_{\bf q}V_{\alpha\gamma}({\bf q})
V_{\gamma\beta}(-{\bf q})\\
&\times \delta(E_{\gamma}({\bf k}-{\bf q})-E_{\alpha}({\bf k})
+\hbar\omega_{\bf q})f_{\gamma}({\bf k}-{\bf q})\Big]\\
+&\textrm{terms with } n_{\bf q}\to n_{-{\bf q}}+1 \textrm{ and } 
\hbar\omega_{\bf q}\to -\hbar\omega_{-{\bf q}}\label{EqJhelp}
\end{split}\end{equation}
where we used that the lower two lines are the complex anti-conjugate of
the upper two lines in the right-hand side of
Eq.~(\ref{EqDenstityMatrix}) after exchanging the indices $\alpha$ and
$\beta$.
Now the completeness of the states $\varphi_{\alpha}(z)$ in the
$z$-part of the Hilbert space provides the relation
\[
\sum_{\alpha\neq\beta}
z_{\beta\alpha}V_{\alpha\delta}({\bf q})=
\sum_{\beta}\left[
\langle \beta |\hat{z} V(z,{\bf q})|\delta\rangle
-z_{\beta\beta}V_{\beta\delta}({\bf q})\right]
\]
to be used in the first summand of Eq.~(\ref{EqJhelp}). In addition
the running index $\delta$ is replaced by $\alpha$. Correspondingly,  
\begin{multline*}
\sum_{\alpha\neq\beta}
V_{\gamma\beta}(-{\bf q})z_{\beta\alpha}=\\
\sum_{\alpha}\left[
\langle \gamma|\hat{z} V(z,-{\bf q})|\alpha\rangle
-z_{\alpha\alpha}V_{\gamma\alpha}(-{\bf q})\right]
\end{multline*}
is used in the second summand with the replacements $\gamma\to \beta$ as 
well as ${\bf k}-{\bf q}\to {\bf k}$ and  ${\bf q}\to -{\bf q}$. 
These operations result in
\begin{equation}\begin{split}
J=&\frac{2\pi e}{\hbar V} \sum_{\alpha\beta,{\bf k},{\bf q}}
n_{-{\bf q}}|V_{\beta\alpha}({\bf q})|^2(z_{\alpha\alpha}-z_{\beta\beta})
\\
&\times 
\delta(E_{\beta}({\bf k})-E_{\gamma}({\bf k}-{\bf q})
+\hbar\omega_{-{\bf q}})f_{\beta}({\bf k})\\
+& \textrm{terms with } n_{\bf q}\to n_{-{\bf q}}+1 \textrm{ and } 
\hbar\omega_{\bf q}\to -\hbar\omega_{-{\bf q}} 
\label{EqJhopping}
\end{split}\end{equation}
which is the standard expression for hopping currents. It can be
interpreted as the sum of scattering transitions from $\beta$ to
$\alpha$, which change the mean 
location of the electron from $z_{\beta\beta}$ to $z_{\alpha\alpha}$.
This is however not the underlying physics, as scattering does not
directly change the particle position. In contrast the entire current
is carried by the polarizations $\rho_{\alpha\beta}$ and
Eq.~(\ref{EqJhopping}) is nothing but an approximation for these
coherences.

\section{Conclusion}

The transport in QCLs and similar structure such as superlattice is
entirely due to coherences, i.e. nondiagonal elements in the density
matrix $\rho_{\alpha\beta}({\bf k})$ if a set of real basis functions
is chosen. The use of NEGF allows for a
spatially and energetically resolved visualization of these coherent 
transport properties.
Neglecting the energetic broadening of the states,
the density matrix equations can be derived from NEGF theory. However, the 
energetic location of the nondiagonal elements is only poorly defined
in this reduction scheme. If the level differences are larger than
the scattering induced broadening $\Gamma$, the
nondiagonal elements of the density matrix are small and can be
approximated by differences in level occupation. In this 
way the frequently used hopping model for the current appears. 
This model suggests the interpretation that the spatial
position of the particles is directly changed by the individual
scattering processes. However one has to keep in mind, that
conventional scattering processes do not change the position of
carrier, but only induce coherences which subsequently drive the current.

\acknowledgments
The author thanks F. Banit, A. Knorr, S.-C. Lee, R. Nelander,
M.F. Pereira, C. Weber, and M. Woerner for detailed discussions 
and long-standing cooperation on the transport theory of QCLs. 
This work was supported by the Swedish Research Council (VR).


\providecommand{\WileyBibTextsc}{}
\let\textsc\WileyBibTextsc
\providecommand{\othercit}{}
\providecommand{\jr}[1]{#1}
\providecommand{\etal}{~et~al.}

\end{document}